# What are suspicious VoIP delays?

Wojciech Mazurczyk, Krzysztof Cabaj, Krzysztof Szczypiorski

Warsaw University of Technology, Faculty of Electronics and Information
Technology, 15/19 Nowowiejska Str., 00-665 Warsaw, Poland
{W.Mazurczyk, K.Cabaj, K.Szczypiorski}@elka.pw.edu.pl

**Abstract**

Voice over IP (VoIP) is unquestionably the most popular real-time service in IP networks today. Recent studies have shown that it is also a suitable carrier for information hiding. Hidden communication may pose security concerns as it can lead to confidential information leakage. In VoIP, RTP (Real-time Transport Protocol) in particular, which provides the means for the successful transport of voice packets through IP networks, is suitable for steganographic purposes. It is characterised by a high packet rate compared to other protocols used in IP telephony, resulting in a potentially high steganographic bandwidth. The modification of an RTP packet stream provides many opportunities for hidden communication as the packets may be delayed, reordered or intentionally lost. In this paper, to enable the detection of steganographic exchanges in VoIP, we examined real RTP traffic traces to answer the questions, what do the "normal" delays in RTP packet streams look like? and, is it possible to detect the use of known RTP steganographic methods based on this knowledge?

## 1. Introduction

Steganography has been used for ages, dating back as far as ancient Greece [19]. Steganographic methods allow for hiding the very existence of the communication, so a third-party observer will not suspect anything if they are unaware of the steganographic exchange. Steganography encompasses information hiding techniques that embed a secret message (steganogram) into the carrier. The carrier is suitable for steganographic purposes if it fulfils two conditions: it is commonly used and the carrier modification caused by the embedding of the steganogram must not be "noticeable" to anyone. The form of the carrier has evolved over time – historical carriers were wax tablets, human skin or letters [19] – now it is instead a digital picture, audio or text.

Recently, a new type of steganography was identified, called *network steganography*. This includes information hiding techniques that utilise, as a carrier, data units and/or their exchange in a telecommunication network. Network steganography can pose a threat to network security, as the current security systems and mechanisms do not provide sufficient countermeasures and are in fact useless against this type of threat. Using steganography for malicious purposes can lead, for example, to confidential information leakage or serve as tools for the distribution of worms and viruses in planning and conducting DDoS (Distributed Denial of Service) attacks [21]. Thus, it is important to answer the question, what real impact may steganographic methods have on network security? The answer may be found through careful evaluation of a particular methods' potential steganographic bandwidth and its possibilities for detection (steganalysis).

VoIP (Voice over IP) is a real-time service which enables users to make phone calls through data networks that use an IP protocol. The popularity of this technology has caused a continuous rise in the volume of VoIP traffic. Thus, it may be increasingly targeted for steganographic purposes, as stated by Lubacz, Mazurczyk and Szczypiorski in [14], and it is therefore important to develop detection methods. To achieve this goal, we must first find an answer to the question, what does an anomaly caused by the use of steganography during a VoIP call look like?

RTP (Real-time Transport Protocol) [22] is the most promising carrier of steganograms in VoIP. RTP provides end-to-end network transport functions suitable for applications transmitting real-time audio. RTP is usually used together with UDP (or rarely TCP) for the transport of digital voice streams. During the conversation phase of the call audio (RTP) streams flow bidirectionally between a caller and a recipient. The rate at which RTP packets flow depends on the codec used, e.g., in the G.711 codec [9], each RTP packet carries 20 ms of voice using 160 bytes; in this case the RTP packet flow rate is 50 packets per second. Thus, even by hiding 1 bit in every RTP packet we gain the quite high steganographic bandwidth of 50 bit/s. In effect, this would allow the user to send about 5 kB of data during a typical VoIP call.

As the authors stated in [16], steganalysis methods must be developed for RTP transmission to enable the detection and/or elimination of hidden communication. To achieve this goal, a steganographic method for inspecting RTP transmission must be developed. Two broad groups of information hiding techniques exist that may affect RTP; the first group is based on modifying the RTP packet header and/or payload, while the second affects the RTP packet stream by modifying the time relation between them. In this study, we focus on the second group of steganographic solutions, because the first is easy to detect and eliminate. Methods for modifying an RTP stream to transmit bits of a steganogram can:

- Affect the sequence order of RTP packets [12] by assigning an agreed-upon order of packets during a predetermined period of time. For example, sending packets in ascending order could indicate a binary one, and descending order a binary zero
- Use different sending rates for the RTP stream [7] – in a simple case, one (the original) rate denotes a binary one, a second rate, achieved, e.g., by delaying RTP packets, means a binary zero
- Modify inter-packet delay [2] – e.g., where predetermined delays between two subsequent RTP packets are used to send single steganogram bit
- Introduce intentional losses [23] by skipping one sequence number while generating RTP packets. Detecting so called "phantom" loss during a predetermined period of time means sending one bit of the steganogram
- Use intentionally delayed packets (by the transmitter) from the RTP stream to carry a steganogram. An example of such a method is LACK (Lost Audio Packets Steganography) [16]. If the delay of the chosen packets at the receiver is considered excessive, the packets are discarded by a receiver not aware of the steganographic procedure. The payload of the intentionally delayed packets is used to transmit secret information to receivers aware of the procedure so no extra packets are generated; for unaware receivers the hidden data is "invisible". More detailed LACK description may be found in [16].

Steganographic methods described above have one common feature – all of them modify delays of the RTP packets. Thus, to evaluate the impact that they may have on network security, real RTP packet delays during VoIP calls should be investigated.

For VoIP, network delays and packets losses have already been thoroughly researched, e.g., in [3], [15] and [5], but not yet in the steganographic context. Moreover, the existing research has focused on measuring overall packet delay and losses rather than their detailed characterisation. Consequently, the main objective of this study was to describe what can happen to packets in an RTP stream while traversing the network based on real VoIP traffic captures. Our research focused on RTP packet delays and all scenarios that may lead to the loss of RTP packets (physically or by the receiver, e.g., jitter buffer). Using this knowledge, we were able to characterise delays that can be introduced into the network and to evaluate the threat which may be posed by steganographic methods that utilise RTP by estimating their steganographic bandwidths. This information will be also needed to develop effective countermeasures.

Thus, the goals of this study were to:

- Characterise the delays and losses for VoIP over the Internet, based on the experiment conducted for an average VoIP call (average duration, connection path length, typical codec, loss concealment method and jitter buffer sizes)
- Identify all scenarios for RTP packet losses, including physical losses and losses caused by jitter buffer (e.g., late packets dropped and buffer overflow), and present the corresponding results
- Evaluate the feasibility of RTP steganographic methods based on real VoIP traffic

The structure of the paper is as follows. Section 2 briefly describes the basics of RTP and the jitter buffer algorithms used in VoIP. In Section 3, the experimental results for VoIP delays are presented and analysed.

Section 4 discusses how the knowledge of real RTP packet delays affects VoIP steganographic methods in use; section 5 concludes our work.

## 2. RTP (Real-time Transport Protocol) packets and VoIP jitter buffers

As mentioned in the introduction, RTP is a crucial protocol for VoIP during the transport of voice packets through IP networks. Usually, RTP packets are generated by the transmitter at a fixed rate, e.g., every 20 ms in the G.711 codec, and they are expected at the receiver at the same rate. However, while traversing the network voice packets may be subjected to such impairments as delay, loss or jitter. Thus, the delays in the received packets can be different from the transmitted ones. This is why there is a need for a receiving buffer, called a jitter buffer. The size of the jitter buffer is crucial for limiting the so-called *mouth-to-ear* delay (which should not exceed 150 ms) and determines the quality of the conversation. If the buffer is too large, the *mouth-to-ear* delay is increased, causing a degradation of call quality. However, if the buffer is too small, overall packet losses are increased due to jitter buffer drops, which can also negatively affect call quality. Thus, the sizing of the jitter buffer always involves a trade-off between increasing the overall delay and minimising losses. Typical jitter buffers for VoIP are sized in the range of 40-80 ms.

Another important fact is associated with the type of jitter buffer used. There are two types of jitter buffers: fixed or adaptive. A fixed buffer has a constant size while an adaptive buffer changes size during the call or between subsequent calls based on information about delays and losses introduced by the network. Adaptive buffers change size in a defined range (e.g., from 40 to 100 ms). Various algorithms for jitter buffering exist and are described, e.g., in [20], [25] or [18]. However, the real problem is that only a few of these proposed algorithms are implemented in practice, and as Wu et al. [26] stated, most popular VoIP applications use fixed-size buffers or adapt to network conditions, but not optimally. In this paper we chose to simulate a simple fixed buffer as specified in [5]. This simple jitter buffer allows for the visualisation of problems that may occur in real RTP streams, and, as mentioned, such fixed-size buffers are still commonly implemented. The jitter buffer operates as follows: after the initiation of a call, before the receiver begins to play back the speech samples to the recipient it continues buffering the RTP packets until the buffer is filled to half capacity. Then, when the next packet above half capacity is received the speech samples are played back.

The next most important mechanism used to limit quality degradation due to packet loss consists of PLC (Packet Loss Concealment) methods. In the simplest scenario, these utilise repetition of the last received packet to substitute for a missing one [9], but more complex algorithms have also been developed [13]. In our implementation, to help preserve voice quality repetition of the last successfully received packet was used to conceal a physically lost or dropped one.

Despite the jitter buffer algorithm and PLC mechanism used in VoIP, packets may be lost; a packet is considered lost if:

- *It is discarded in the network* (Fig. 1, point b) – in this case it never reaches the receiver. Such a situation may be caused, e.g., by buffer overflow in some intermediate device caused by a bottleneck within a network. We refer to such losses as *physical losses.*
- *It is dropped by the jitter buffer* (Fig. 1, point c) – when an RTP packet is excessively delayed due to network latency it reaches the receiver but is useless as it cannot be used for voice reconstruction; thus, it is discarded and counted as lost. Moreover, due to so-called delay spikes, the jitter buffer, in addition to dropping late packets (drops caused by *buffer underflow*) may also drop subsequent RTP packets because they may all arrive simultaneously and the size of the jitter buffer may be insufficient to store them all (*buffer overflow*).

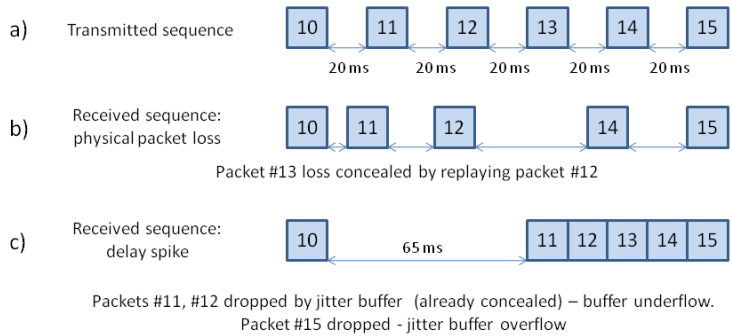

**Fig. 1** Packet losses in VoIP.

VoIP statistics regarding losses should distinguish between physical losses and losses caused by jitter buffer. Moreover, we need to know what realistic VoIP inter-packet delays are and what they look like. Do the delays and losses happen singly or in series, and if so, can these series be characterised? Another important consideration is whether any method may utilise an intentional reordering of RTP packets – is this realistic for IP telephony in today's Internet? We address these and other questions in the next sections of this paper.

## 3. What can actually happen to RTP packets while traversing a network?

For a practical evaluation of the feasibility of steganographic methods utilising RTP as specified in the introduction, we assumed that the VoIP endpoints exchanging the RTP streams are also the sender and the receiver of the steganogram (but this is not the only possible scenario, as stated by Mazurczyk and Szczypiorski in [16]).

To evaluate the real VoIP delays of RTP packets, an experiment was conducted. Calls were established from Warsaw, Poland to Cambridge, UK (see Fig. 2) through the Internet using the very popular free SIP-based softphone *X-lite* [27] and an SIP proxy server (located in Warsaw). The distance between the cities is ~1,800 km. One hundred calls were captured using *Wireshark* (www.wireshark.org) between 27 October and 4 November, 2009 during working hours, which resulted in total number of 2,825,076 packets transmitted. The communication path between the cities represents an average Internet connection path of about 16 hops [1, 28, 6]. Audio was coded with the ITU-T G.711 A-law PCM codec (20 ms of voice per packet, 160 bytes of payload). The average call duration for the experiment was chosen based on the average duration of calls using Skype [24] and other VoIP service providers. In [8], the results obtained show that the average call duration on Skype was about 13 minutes, while for other VoIP providers it is typically between 7 and 11 minutes. Thus, we chose an experimental call duration of nine minutes.

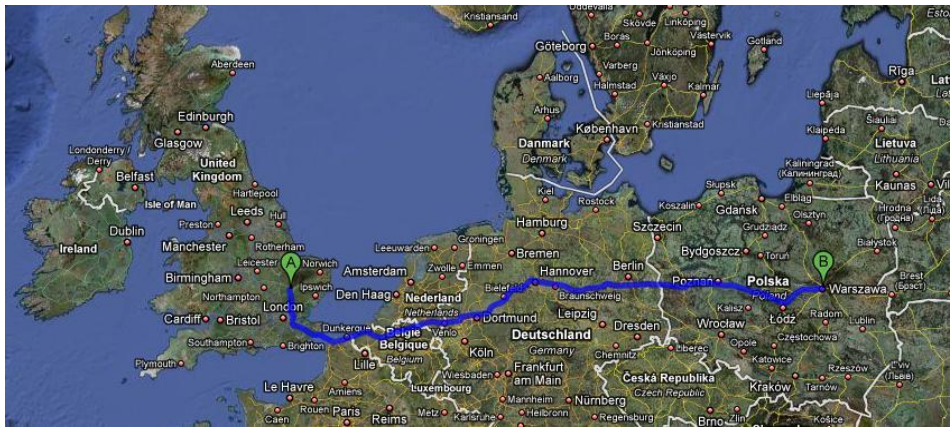

**Fig. 2** VoIP experimental evaluation scenario – calls from Cambridge, UK (A) to Warsaw, Poland (B) (http://maps.google.com).

In the experiment, we developed custom software to analyse delays and losses occurring in the RTP stream of the captured call traces and to simulate different fixed buffer sizes to be able to evaluate the relationships between RTP packet delays, losses and jitter buffer size. The jitter buffer sizes used and the number of buffered packets after which the playback of the voice samples began are specified in Table 1. The description of the jitter buffer algorithm was described in detail in the previous section. For each experimental call we measured packets dropped by the jitter buffer, delayed packets and physical losses.

| Jitter buffer size [ms] | 20 | 40 | 60 | 80 | 100 | 120 |
|---|---|---|---|---|---|---|
| No. of initially buffered packets | 1 | 1 | 2 | 2 | 3 | 3 |
| Playback starts after receiving [packets] | 2 | 2 | 3 | 3 | 4 | 4 |

Table 1. Jitter buffer characteristic parameters.

For each call quality was assessed using the ITU-T E-model [11], which is a quality objective assessment method for transmission planning. The E-model expresses call quality as an *R* factor which ranges in value from 0 (worst quality) to 100 (best quality). For real VoIP traffic, Cole and Rosenbluth [5] proposed a simplified formula for *R* calculation based on VoIP performance monitoring, which takes into account only impairments caused by losses and delays, as follows:

$$R = 94.2 - I_d - I_{ef} \qquad (1)$$

where $I_d$ denotes impairments caused by delays and $I_{ef}$ impairments caused by losses. $I_d$ was calculated based on mouth-to-ear delay (*d*) as proposed in [5]:

$$I_d = 0.024 + 0.11 \cdot (d - 177.3) \, H(d - 177.3) \qquad (2)$$

where *H(x)* is the Heaviside (or step) function defined as:

$$H(x) = \begin{cases} 0 & \text{if } x < 0 \\ 1 & \text{if } x \geq 0 \end{cases} \qquad (3)$$

$I_{ef}$ was calculated based on an equation given by [5] that was derived explicitly for the G.711 codec, additionally concerning random losses:

$$I_{ef} = 30 \cdot \ln(1 + 15 \, p_L) \qquad (4)$$

where $p_L$ denotes the probability of RTP packet loss.

Based on the E-model and results from our experiments, an *R* factor was obtained. This was then converted into an MOS score ranging from 1 (bad quality) to 5 (good quality) [10], which is typically used for expressing the quality of VoIP calls, using the known formula:

$$MOS = 1 + 0.035 \cdot R + 7 \cdot 10^{-6} \cdot R(R-60)(100-R) \qquad (5)$$

For the experimental scenario and assumptions presented above the following results were obtained.

Call quality results are presented in Fig. 3. It is often assumed that an MOS score equal to or greater than 3.6 is considered to be of comparable quality to traditional PSTN (Public Switched Telephone Network) calls [3]. By this standard, the quality of the experimental calls using the 80-, 100- and 120-ms buffer sizes can be judged as good, as less than ~20% of these calls were of a quality lower than 3.6. For the 60-ms jitter buffer about 30% of the calls were of lower quality than 3.6, while for the 20-ms jitter buffer it was about 75% of all calls.

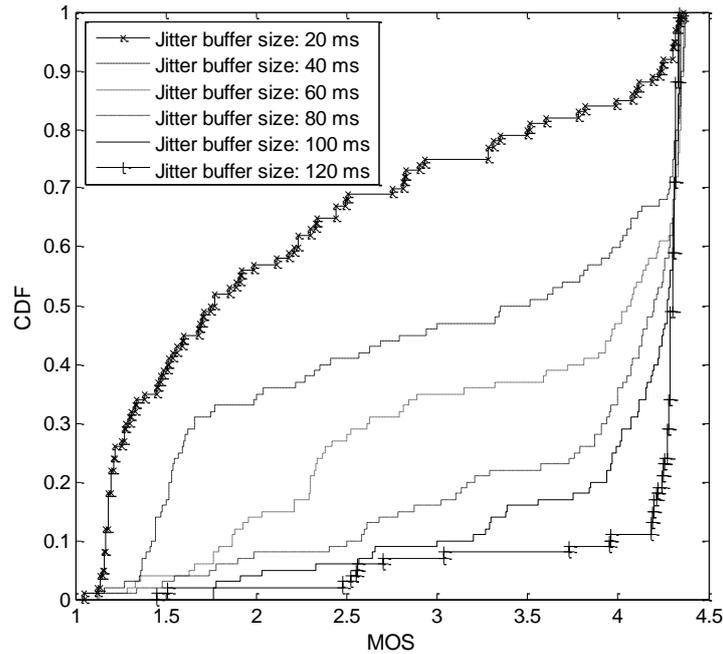

**Fig. 3** CDF of MOS scores for different jitter buffer sizes.

It was of interest to plot the cumulative distributions of the two most important impairments in the experimental VoIP data: delay and loss. The results for physical packet losses are presented in Fig. 4.

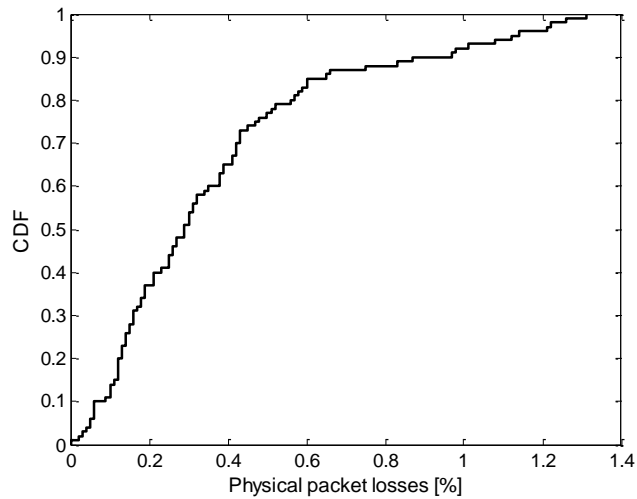

**Fig. 4** CDF of physical packet losses for the experimental data.

The above figure illustrates that losses caused by the network (packets that never reach the receiver) did not exceeded 0.5% of all RTP packets in more than 80% of the calls. These results somehow confirms earlier research in that area. For example, Borella et al. [4] analysed a month of Internet packet loss statistics for speech transmission and their findings are that physical packet losses for the three paths, all in the U.S., ranged from 0.4% to 3.5%.

What we want to explore next is the RTP inter-packet delays. Here, we wanted to know how many packets arrived late if we assumed a certain delay threshold. For our experiment, we chose threshold values from 20 to 100 ms with a step of 20 ms. The results obtained are presented in Fig. 5.

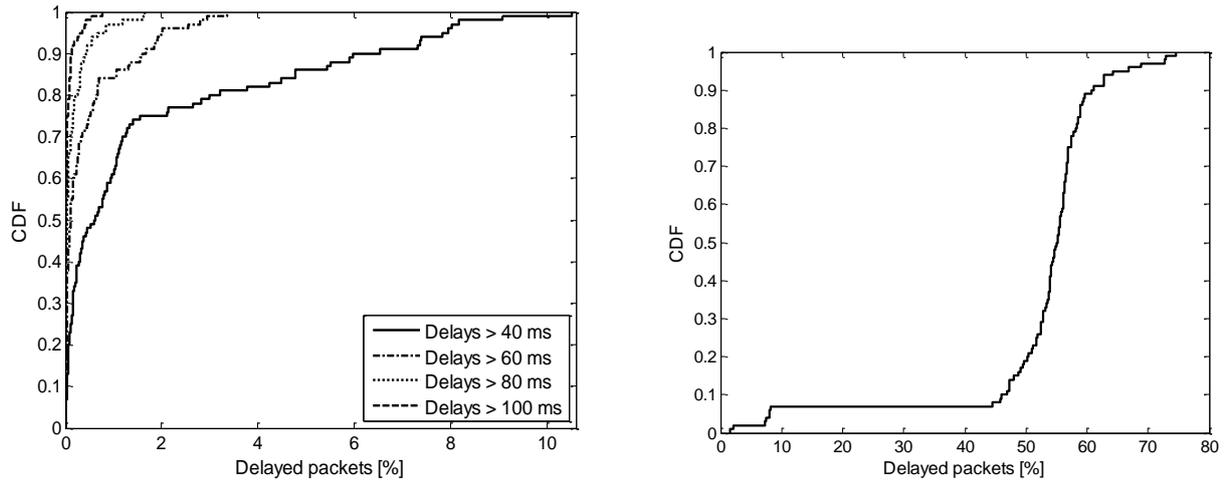

**Fig. 5** CDF of RTP packet delays for different delays thresholds – 40, 60, 80, 100 ms (left), 20 ms (right).

From the figures above it can be seen that there was a great difference in the number of delayed RTP packets between a packet delay of 20 ms and the remaining delay values. This was caused by the packet generation time interval in the transmitter – packets were sent each 20 ms. Thus, if there was any delay, even the slightest, in a packet's reception introduced by the network or by clock skew it was counted as delayed. It is obvious that the larger the assumed delay threshold the lower the number of delayed packets. For example, about 30% of the calls experienced 2% or more packets delayed more than 40 ms, while only about 5% of all calls had about 1% or more packets delayed by more than 80 ms.

Next, we compared how many of the delayed packets presented in the figure above resulted in losses caused by jitter buffer drop. First, in Fig. 6 we present packet drops caused by jitter buffer. As expected, with an increase in the size of the jitter buffer the number of packets dropped decreased. For example, more than 40% of the packets were dropped in 35% of the calls using a 40-ms jitter buffer, whereas it was about 5% of all VoIP calls for the 80-ms jitter buffer.

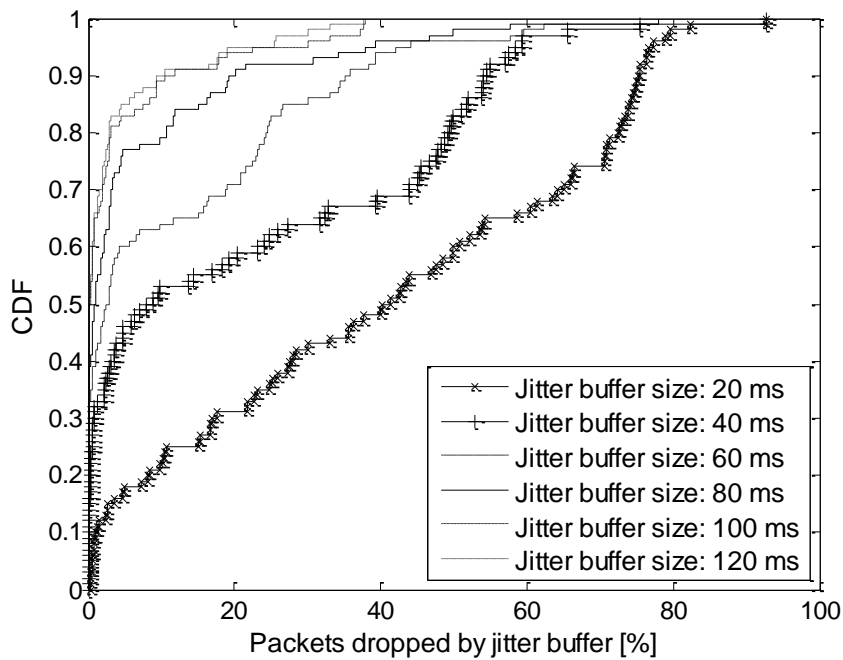

**Fig. 6** CDF of RTP packet drops by jitter buffer for different buffer sizes.

The results concerning the influence that delayed RTP packets have on packet losses caused by a too-small jitter buffer are presented in Fig. 7 (for 40-ms and 80-ms jitter buffers).

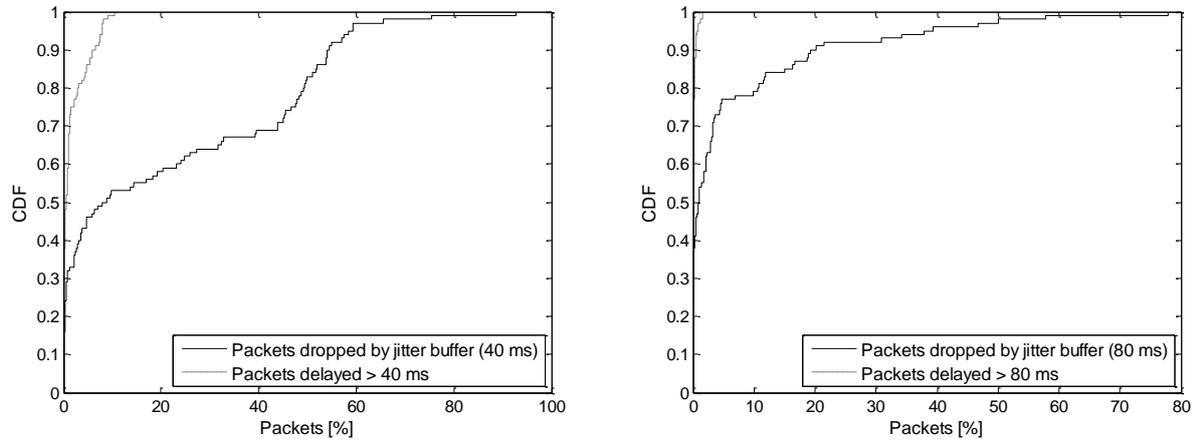

**Fig. 7** Relationship between delayed and dropped packets for jitter buffers of 40 ms (left) and 80 ms (right).

With the 40-ms jitter buffer almost 50% of all calls experienced ~10% or more packet drop, while ~10% or more of the packets were delayed by more than 40 ms for only 10% of the calls. For 80-ms jitter buffer, ~ 10% or more buffer drops were observed for about 25% of the calls, with only a small number of the packets delayed by more than 80 ms. Thus, the larger jitter buffer yielded a greater number of delayed packets that were not lost and could be used for voice reconstruction. Simultaneously, a larger buffer adds more delay to the conversation, which may affect call quality if *mouth-to-ear* delay exceeds 150 ms. Total losses, including packets dropped by jitter buffer and physical losses, are presented in Fig. 9.

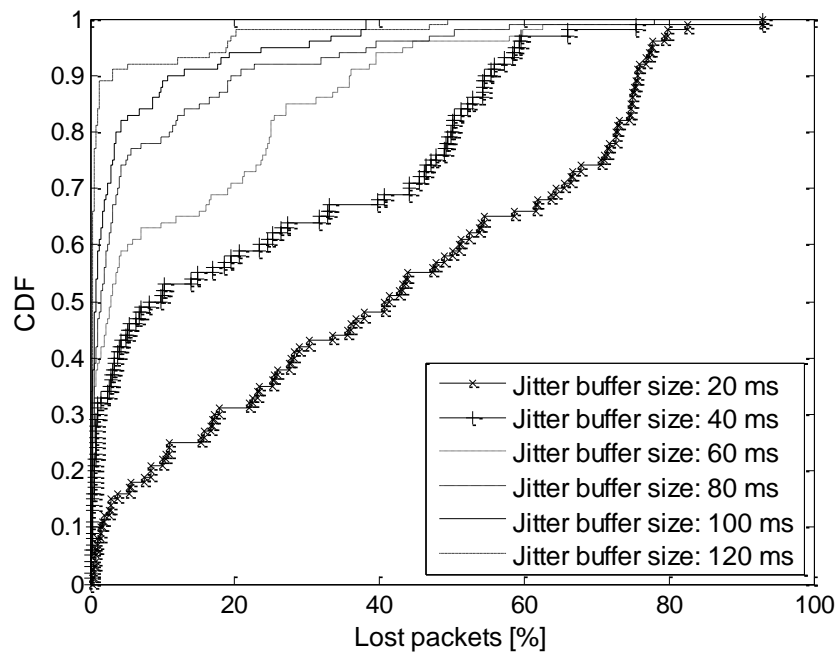

**Fig. 9** Total packets lost due to physical losses and drops by the jitter buffer.

For the G.711 codec, which has PLC (Packet Loss Concealment) functionality, the maximum tolerable packet loss is 5% [17]. Thus, for our experimental data it would be best to use an 80-ms or larger jitter buffer to preserve conversation quality as for this size almost 80% of the calls experienced losses lower than 5%.

Next, we compared physical losses and losses caused by jitter buffer (Fig. 10).

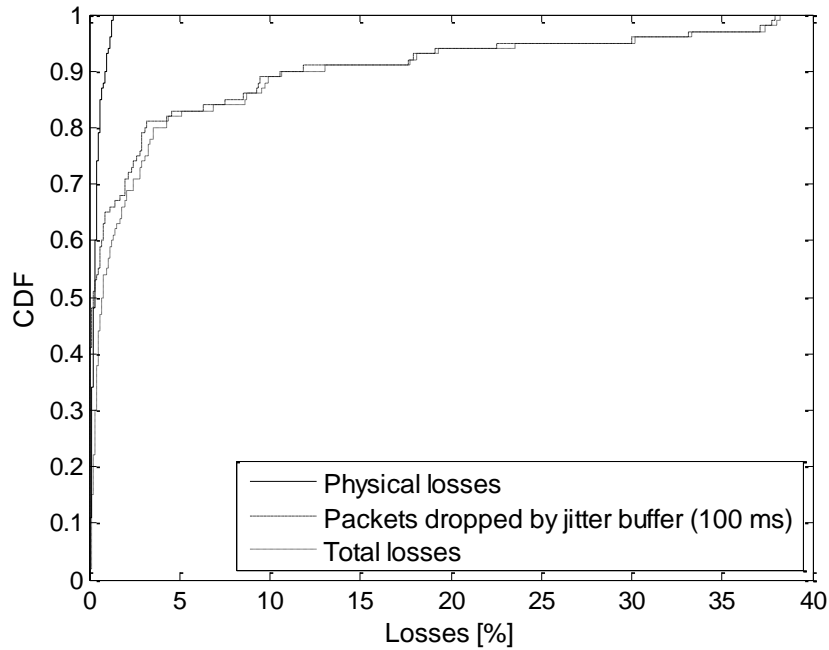

**Fig. 10** Relationship between physical losses and losses caused by jitter buffer (100 ms).

The fraction of the physical losses was so small that the main factors determining total losses were those associated with buffer drops. As stated in the introduction, there are two types of jitter buffer losses: drops caused by jitter buffer overflow and those which are caused by late packets. The buffer overflows when it is full and the next received packet cannot be stored, thus it must be discarded (for logging purposes we noted this event as a D1 drop). The second type of jitter buffer drop is caused by RTP packets which are received too late, so they are not present at the receiver when they should be played to the recipient. In this case the PLC mechanism fills the gaps by replaying the last successfully received packet. When these packets finally reach the receiver, they cannot be used for voice reconstruction, and they are dropped as a result (for logging purposes we noted this event as a D2 drop). In both cases described, the dropped RTP packets physically reach the receiver but they are discarded by the jitter buffer and never used for voice reconstruction.

Late packet drops (D2) are usually caused by delay spikes, i.e., at some point in the connection there is a great increase in inter-packet delay which results in buffer underflow, and the late packets are not needed for voice reconstruction (because they have already been concealed) and are dropped. However, jitter buffer overflows (D1) happened more frequently for calls that experienced the following event: at the beginning of the call a burst of RTP packets came nearly simultaneously (i.e., the inter-packet delay was about zero). Such an event, especially for small jitter buffers, results in buffer overflow and influences the rest of the conversation as well by introducing subsequent drops whenever the inter-packet delay differs, even slightly, from the RTP packet generation time (20 ms).

The results from the application of the software tool developed for RTP stream analysis revealed two often-observed sequences of events that produced high levels of packet drop. Both situations are presented in Table 2, which contains sample sections of two logs from the developed research software. Here, only the sequence number of the RTP packets (the number after the *seq* string) and the event type (the string after the colon) are considered in the given excerpts; the two other numbers represent analysed RTP packet numbers from the beginning of the given RTP stream and from the beginning of the recorded *Wireshark* file (these values were used for testing and debugging purposes).

```
900[1861,seq 8101],10: D1      37[101,seq 5331],20: U
905[1871,seq 8101],10: R       37[101,seq 5328],20: D2
905[1871,seq 8106],10: D1      38[103,seq 5332],20: U
910[1881,seq 8106],10: R       38[103,seq 5329],20: D2
910[1881,seq 8111],10: D1      39[105,seq 5333],20: U
915[1891,seq 8111],10: R       39[105,seq 5330],20: D2
915[1891,seq 8116],10: D1      40[107,seq 5334],20: U
920[1901,seq 8116],10: R       40[107,seq 5331],20: D2
```

**Table. 2** Sample logs from the research software developed for RTP stream analysis (U refers to a buffer underflow event and R represents invocation of the PLC mechanism; the sequence number from RTP header is given after the *seq* string).

The left column in Table 2 presents drops due to buffer overflows (D1). Here, a received RTP packet is dropped due to the full jitter buffer. In effect, the PLC mechanism had to subsequently reconstruct the packet dropped earlier, leading to an R event. This situation can be observed, for example, in the first two lines, concerning the RTP packet with sequence number 8101.

The right column of Table 2 presents the second type of jitter buffer drops (D2). In contrast to the previous situation, in this case drops are associated with buffer underflow events (U). Because the jitter buffer is empty the PLC mechanism had to reconstruct a packet, and as a result a U event appears in log. Later, when the original late packet arrived, due to the previous reconstruction it had to be dropped. This sequence, concerning the packet with sequence number 5331, can be observed in first and last lines.

Not surprisingly, for both types of jitter buffer drops increasing the jitter buffer size caused a decrease in the total buffer losses. It should be also noted that drops caused by buffer overflows were more rapidly compensated for with increased buffer size (see Figs. 12 and 13). For smaller jitter buffer sizes, i.e., from 20 to 60 ms, losses due to jitter buffer overflows dominated, while for buffers larger than 60 ms losses caused by late packets took precedence (Fig. 11).

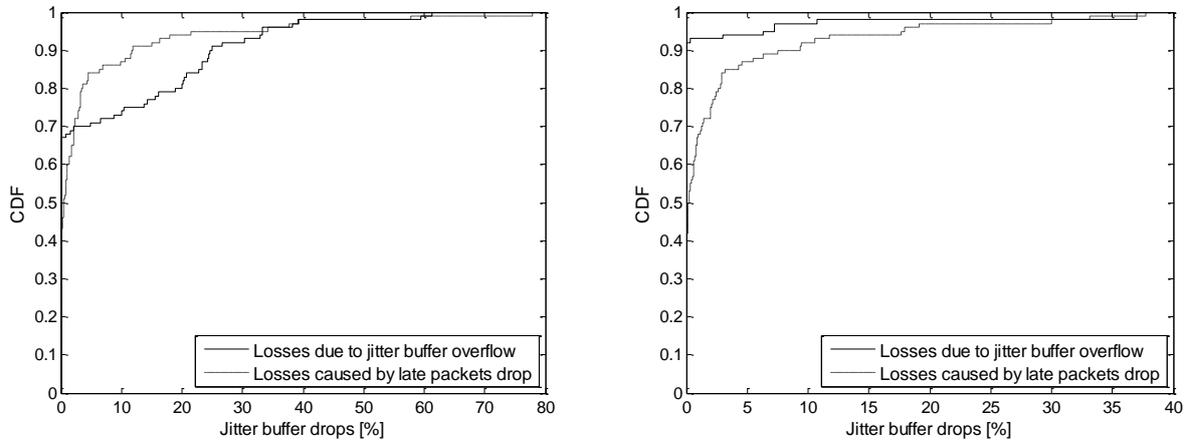

**Fig. 11** Comparison of different types of jitter buffer drops for 60-ms (left) and 100-ms (right) jitter buffers.

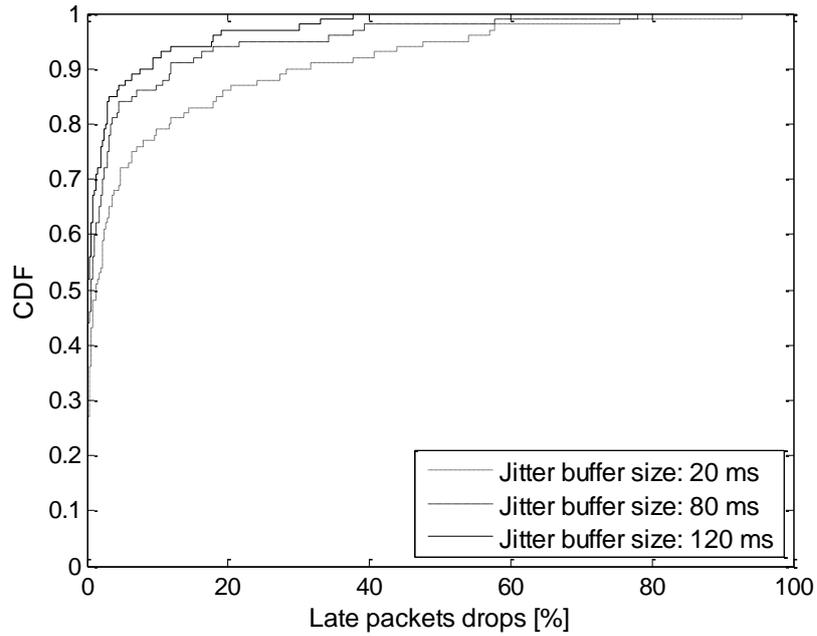

**Fig. 12** Late packet drops by jitter buffers (20-, 80- and 120-ms).

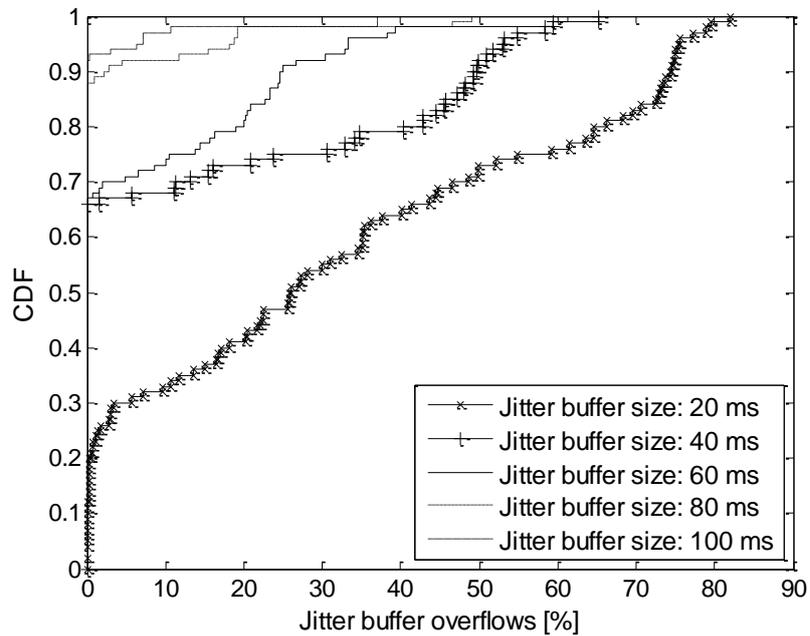

**Fig. 13** Packets dropped due to jitter buffer overflow.

Because the jitter buffers sized at 20, 40 and 60, 80 and 100, and 120 ms start playing the voice samples after buffering the same number of RTP packets (1, 2 and 3, respectively), the curves representing losses due to late packet drops for these buffers were the same (hence, overlapping curves are not presented in Fig. 12).

We also observed that a large number of the experimental calls followed a pattern of only a single type of jitter buffer drop, i.e., if there were a lot of drops caused by buffer overflows, the level of late packet drops for the same call was rather low and *vice versa*.

Finally, it must be emphasised that during the performed experimental calls there *were no reordered RTP packets*. This means that while delays, even high delays, are possible for the RTP packets they do not lead to their reordering.

## 4. Feasibility of RTP steganographic methods based on real VoIP traffic

First, let us consider steganographic methods that affect the sequence of RTP packets. For a sequence of $n$ RTP packets, the potential number of steganogram bits is $log_2(n!)$; thus, the steganographic bandwidth ($S_B$) may be expressed as:

$$S_B = \frac{i \cdot \log_2 n!}{T} \quad [bits/s] \qquad (6)$$

where $T$ denotes VoIP call duration (in seconds) and $i$ is the number of time intervals in which a steganogram will be detected. For example, if we assume that we try to send a steganogram using a sequence of 10 subsequent RTP packets (for G.711 it is interval of 0.2s, so $i$=2700), for the same call duration as the experimental ones (540 s) we achieve a steganographic bandwidth of about 100 bits/s. However, it must be noted that, as mentioned above, there were *no reordered RTP packets*, so applying such a method will be trivial to detect. Moreover, affecting the sequence of the RTP packets may lead to a deterioration of conversation quality as the jitter buffer may be unable to compensate for intentional packet reordering.

Next, let us consider steganographic methods that utilise different RTP packet-sending rates. In the simplest case, the original generation rate of the RTP packets denotes sending a binary one and second rate is achieved, e.g., by delaying RTP packets, which means sending a binary zero. If $h$ different methods of sending RTP packets are used it is possible to send $log_2 h$ bits of a steganogram. This may be expressed as:

$$S_B = \frac{i \cdot \log_2 h}{T} \quad [bits/s] \qquad (7)$$

For example, if $h = 2$ and we assume a VoIP call duration of nine minutes and a steganogram is sent each second then we achieve a steganographic bandwidth of about 1 bit/s. A similar method is based on modifying RTP inter-packet delay, where predetermined delays between two subsequent RTP packets are used to send one steganogram bit.

For these two methods let us consider Fig. 11, showing inter-packet delay diagrams for two experimental calls that were chosen based on different delay statistics. The left diagram in Fig. 11 presents an experimental call that experienced high inter-packet delays during the call and the right diagram shows the opposite situation.

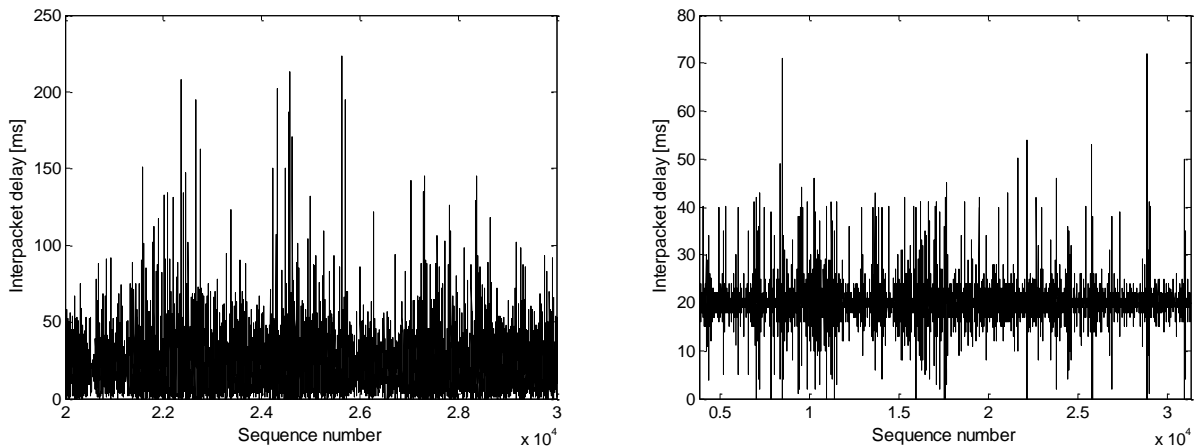

**Fig. 11** High (left) and low (right) inter-packet delays of two selected experimental calls.

Note that the difference in inter-packet delay for these two diagrams is quite high and the delay spikes are distributed rather randomly. If we now assume a low inter-packet delay and if we apply the steganographic method utilising two different rates of RTP packet generation, the resulting diagram, analogous to those presented above, will be similar to that presented in Fig. 12.

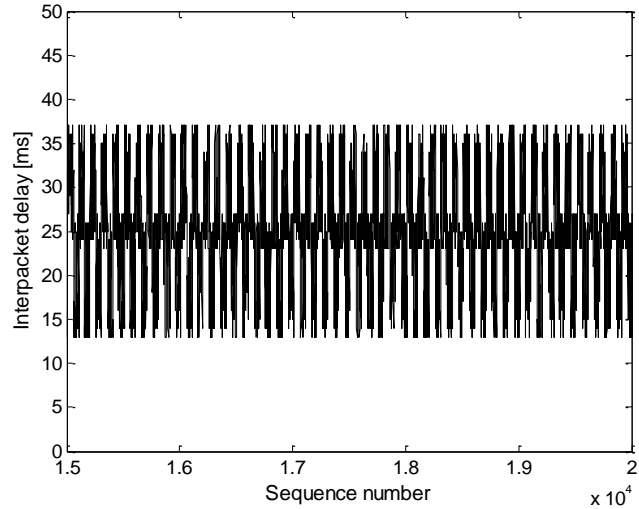

**Fig. 12** Exemplary inter-packet delays for a steganographic method utilising two different rates for RTP packets.

If the RTP packet generation rate is intentionally modified in order to send a steganogram, a certain regularity in inter-packet delays may be observed. Thus, the detection of such method is easy. Moreover, if the RTP packets experience high inter-packet delays (Fig. 11, left diagram) then reception of the steganogram bits may be difficult. The same argument applies to the steganographic method that modifies inter-packet delays.

Let us focus on intentional losses by skipping one sequence number while generating RTP packets. Detecting such so-called "phantom" loss during a predetermined time period means sending one bit of the steganogram. As in the case of the method which modified inter-packet delays, the reception of the steganogram bits may be disrupted due to losses introduced by the network. For the experimental data the average packet loss was 0.37% (about 100 packets), which would make detection of steganogram bits difficult.

Moreover, from a practical point of view such a method is characterised by a rather low steganographic bandwidth, which may be expressed as:

$$S_B = \frac{i}{T} \quad [bits/s] \qquad (8)$$

For example, if we assume that intentional losses will be invoked every five seconds during the call, the steganographic bandwidth will be about 0.2 bits/s.

Now, consider a method that uses intentionally delayed packets in the transmitter of the RTP stream to carry a steganogram such as LACK (Lost Audio Packets Steganography). As proven by the results discussed in Section 3, LACK can utilise both types of events which lead to packet dropping by jitter buffer (D1 and D2): delay spikes (by intentionally increasing inter-packet delay) and an RTP-packet burst at the beginning of the call, which can cause buffer underflows during the remaining part of the connection. The late packet drops (D2) occur almost twice as often as drops due to buffer overflows (D1), so it would be easy to explain that the probability of a packet being late is greater than its probability of arriving too soon. In a typical, nonsteganographic VoIP call, such events happen often enough to provide quite a good steganographic bandwidth, which can be expressed as:

$$S_B = r \cdot p_L \ [bits/s] \qquad (9)$$

where *r* denotes the codec output rate (e.g., 64 kbit/s for G.711) and $p_L$ is the probability of intentional RTP packet loss introduced by LACK. For example, if the G.711 codec is used and there is a 1% intentional loss the steganographic bandwidth achieved is about 640 bits/s.

Let us consider that a 100-ms jitter buffer is the size for which an acceptable voice quality was achieved (see section 3). The average number of drops due to buffer overflow would then be about 300 during the whole connection (with a standard deviation of 1,490). Assuming that during the call about 150 intentionally invoked drops are introduced, the potential steganographic bandwidth achieved is about 350 bit/s. The average number of drops caused by delay spikes during the connection is about 750 (with a standard deviation of 1,882), resulting in steganographic bandwidth of about 900 bit/s if half of the average drops are invoked intentionally. Moreover, we can utilise a combine of these two types of drops during the same connection which results in an increased steganographic bandwidth. Because of the high standard deviations, this could be interpreted as making it extremely hard to predict the number of drops, thus the detection of LACK is not easy but is also very crucial. Of course, introducing jitter buffer losses must be carefully controlled to minimise the chance of detecting inserted data and to avoid excessive deterioration of voice quality. Additionally, packet losses introduced by the network must be carefully monitored. Because LACK uses legitimate RTP traffic, it thus increases overall packet losses. To ensure that the total packet loss introduced by the network and by LACK will not degrade the perceived quality of the conversation, the level of packet loss used for steganographic purposes must be controlled and dynamically adapted.

The high, potentially steganographic bandwidth of LACK makes it the most dangerous method among all those presented in this study that may influence an RTP stream. Thus, developing and implementing steganalysis methods for LACK is crucial.

## 5. Conclusions and future work

In this study delays and losses of voice (RTP) packets during real VoIP traffic were inspected in detail. Modifying the RTP packet stream potentially provides many of opportunities for hidden communication, as the packets may be delayed, reordered or intentionally lost. To assess whether RTP streams are suitable for steganographic purposes, an experiments was conducted, in which 100 average VoIP calls (of typical duration, connection path length, codec, loss concealment method and jitter buffer sizes) were performed. The experimental data was evaluated with respect to RTP packet losses including physical losses and losses caused by jitter buffer, where late packet drops and buffer overflows were distinguished, and the corresponding results for such losses were presented. Most importantly, the results were analysed to evaluate the feasibility of implementing RTP steganographic methods based on real VoIP traffic.

Steganographic traffic is harder to detect, when its characteristic is similar to normal (innocent) traffic that can be observed in a network. The results obtained proved that some of the proposed methods may be quite easily detected, as, e.g., reordering was not present in the captured data, thus the feasibility of such methods is questionable. On the other hand, when steganographic method mimics some often-observed behaviour of the protocol, its detection may be hard. For example, LACK may mimic delay spikes, characteristic formation of packets which can lead to packet drops at the receiving end. In result, this method is quite feasible, and thus it may be considered as a threat to network security. LACK can use RTP packet sequences that will surely lead to jitter buffer losses by causing late packet drops or jitter buffer overflows. LACK may provide a potential steganographic bandwidth of hundreds of bits per second and be more difficult to detect than the other steganographic methods considered here. Further research concerning analysing VoIP traffic should identify often-observed protocols behaviours (packet exchanges) that can be utilized by potential new steganographic methods. Usage of such methods can lead to hiding of steganographic data that may be even more difficult to detect.

In future work, more VoIP data must be analysed to verify and confirm with greater accuracy the results obtained and presented in this paper. Moreover, it was shown that some steganographic methods utilising RTP can pose a serious threat to network security, hence detection solutions must be designed and developed.

**References**

1. Begtasevic F, Van Mieghem P (2001) Measurements of the hopcount in Internet. In: Proc. of the Passive and Active Measurement. 2001
2. Berk V, Giani A, Cybenko G (2005) Detection of Covert Channel Encoding in Network Packet Delays, Tech. Rep. TR2005-536, Department of Computer Science, Dartmouth College, November 2005, URL: http://www.ists.dartmouth.edu/library/149.pdf
3. Birke R, Mellia M, Petracca M, Rossi D (2007) Understanding VoIP from Backbone Measurements, In Proc. of 26th IEEE International Conference on Computer Communications (INFOCOM 2007), May 2007, pp. 2027-35, ISBN 1-4244-1047-9
4. Borella M, Swider D, Uludag S, Brewster G (1998) Internet packet loss: Measurements and implications for End-to-End QoS, In Proc. of International Conference on Parallel Processing, August 1998
5. Cole R G, Rosenbluth J H (2001) Voice over IP performance monitoring, ACM SIGCOMM Computer Communication Review, vol. 31 no. 2, pp. 9-24, April 2001
6. Fei A, Pei G, Liu R, Zhang L (1998) Measurements on Delay and Hop-Count of the Internet, Proc. IEEE GLOBECOM'98, 1998
7. Girling C G (1987) Covert Channels in LAN's, IEEE Trans. Software Engineering, vol. SE-13, no. 2, February 1987, pp. 292–96
8. Guha S, Daswani N, Jain R (2006) An Experimental Study of the Skype Peer-to-Peer VoIP System, Sixth International Workshop on Peer-to-Peer Systems (IPTPS), February 2006
9. ITU-T Recommendation: G.711 (1988) Pulse code modulation (PCM) of voice frequencies, November 1988
10. ITU-T Recommendation: P.800 (1996) Methods for subjective determination of transmission quality, September 1996
11. ITU-T, Recommendation G.107 (2002) The E-Model: a computational model for use in transmission planning, 2002
12. Kundur D, Ahsan K (2003) Practical Internet Steganography: Data Hiding in IP, Proceedings of the Texas Workshop on Security of Information Systems, April 2003
13. Liang Y J, Farber N, Girod B (2003) Adaptive playout scheduling and loss concealment for voice communications over IP networks, IEEE Transactions on Multimedia, vol. 5, no. 4, 2003, pp. 532-543
14. Lubacz J, Mazurczyk W, Szczypiorski K, Vice over IP, In: IEEE Spectrum, ISSN: 0018-9235, February 2010, pp. 40-45
15. Markopoulou A P, Tobagi F A, Karam M J (2002) Assessment of VoIP Quality over Internet Backbones, IEEE Infocom, New York, NY, June 2002
16. Mazurczyk W, Szczypiorski K (2008) Steganography of VoIP Streams, In: R. Meersman and Z. Tari (Eds.): OTM 2008, Part II - Lecture Notes in Computer Science (LNCS) 5332, Springer-Verlag Berlin Heidelberg, Proc. of The 3rd International Symposium on Information Security (IS'08), Monterrey, Mexico, November 2008, pp. 1001-1018
17. Na S, Yoo S (2002) Allowable Propagation Delay for VoIP Calls of Acceptable Quality. In: Chang, W. (ed.) AISA 2002. LNCS, vol. 2402, pp. 469–480. Springer, Heidelberg, 2002
18. Narbutt M, Murphy L (2003) VoIP playout buffer adjustment using adaptive estimation of network delays. In Proceedings of 18th International Teletraffic Congress (ITC-18), 2003, pp. 1171–1180
19. Petitcolas F, Anderson R, Kuhn M (1999) Information Hiding – A Survey IEEE, Special Issue on Protection of Multimedia Content, July 1999
20. Ramjee R, Kurose J, Towsley D, Schulzrinne H. (1994) Adaptive playout mechanisms for packetized audio applications in wide-area networks. In Proceedings of the IEEE INFOCOM 1994, 1994, pp. 680–688
21. Schechter S E, Smith M D (2003) Access for sale, ACM Workshop on Rapid Malcode (WORM'03), ACM SIGSAC, November 2003
22. Schulzrinne H, Caspkner S, Frederick R, Jacobson V (2003), RTP: A Transport Protocol for Real-Time Applications, IETF, RFC 3550, July 2003

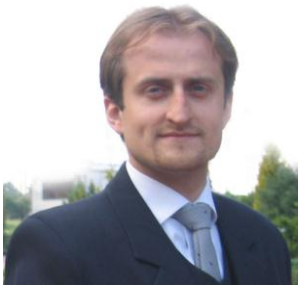

Wojciech Mazurczyk holds an M.Sc. (2004) and a Ph.D. (2009) in telecommunications from the Faculty of Electronics and Information Technology, Warsaw University of Technology (WUT, Poland) and is now an Assistant Professor at WUT and the author of over 40 scientific papers and over 25 invited talks on information security and telecommunications. His main research interests are information hiding techniques, network security and multimedia services, and he is also a research leader of the Network Security Group at WUT (secgroup.pl). Personal website: http://mazurczyk.com.

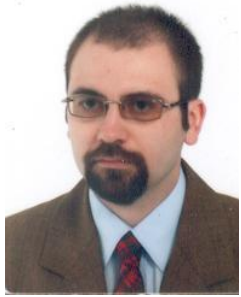

Krzysztof Cabaj holds an M.Sc (2004) and a Ph.D. (2009) in computer science from the Faculty of Electronics and Information Technology, Warsaw University of Technology (WUT), and is an Assistant Professor at WUT and a researcher in the Network Security Group formed at WUT. He has served as an Instructor of Cisco Academy courses: CCNA, CCNP and NS at the International Telecommunication Union Internet Training Centre (ITU-ITC). His research interests include network security, honeypots and data-mining techniques. He is the author or co-author of over 20 publications.

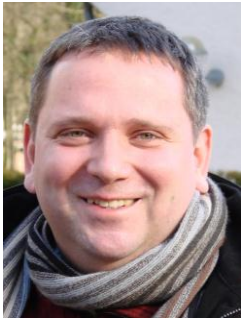

Krzysztof Szczypiorski holds an M.Sc. (1997) and a Ph.D. (2007) in telecommunications both with honours from the Faculty of Electronics and Information Technology, Warsaw University of Technology (WUT), and is an Assistant Professor at WUT. He is the founder and head of the International Telecommunication Union Internet Training Centre (ITU-ITC), established in 2003. He is also a research leader of the Network Security Group at WUT (secgroup.pl). His research interests include network security, steganography and wireless networks. He is the author or co-author of over 110 publications including 65 papers, two patent applications, and 35 invited talks.